\documentclass[12pt,reqno]{amsart}

\usepackage{fullpage}
\usepackage{bm}
\usepackage{amsmath,amssymb,amsthm}
\usepackage{graphicx}
\usepackage{subfigure}

\begin{document}

\newtheorem{thm}{Theorem}
\newtheorem{cor}{Corollary}
\newtheorem{lem}{Lemma}
\newtheorem{prop}{Proposition}
\def\Ref#1{Ref.~\cite{#1}}

\def\const{\text{const.}}
\def\Rnum{{\mathbb R}}
\def\sgn{{\rm sgn}}

\def\t{\mathrm{t}}

\def\eff{\text{eff.}}
\def\peak{\text{peak}}
\def\lump{\text{sphal.}}
\def\solitary{\text{solitary}}

\def\sech{{\rm sech}}
\def\coth{{\rm coth}}
\def\csch{{\rm csch}}
\def\arctanh{{\rm arctanh}}
\def\arccosh{{\rm arccosh}}

\tolerance=80000
\allowdisplaybreaks[4]

\title{Spectral problem for instability of sphalerons\\ in quartic Klein-Gordon models}

\author{
Stephen C. Anco${}^1$\\
\\\\
D\lowercase{\scshape{epartment}} \lowercase{\scshape{of}} M\lowercase{\scshape{athematics and}} S\lowercase{\scshape{tatistics}}\\
B\lowercase{\scshape{rock}} U\lowercase{\scshape{niversity}}\\
S\lowercase{\scshape{t.}} C\lowercase{\scshape{atharines}}, C\lowercase{\scshape{anada}}
}

\thanks{${}^1$ sanco@brocku.ca }

\begin{abstract}
In the energy functional of nonlinear Klein-Gordon models,
sphalerons arise from a saddle point between true and false vacua. 
The spectral problem for the (in)stabilty of sphalerons
in quartic models is shown be governed by a Heun differential equation,
which has three finite regular singular points,
after a certain change of variable. 
This allows an explicit formulation of the eigenfunctions and eigenvalues
to be obtained in terms of local Heun functions.
These functions are generalizations of hypergeometric functions
and possess Frobenius series representations. 
Good approximations are found for certain ranges of the parameter in the potential. 
\end{abstract}

\maketitle

\section{Introduction}

Nonlinear Klein-Gordon models are of widespread interest in
mathematics and physics.
Various kinds of nonlinear phenomena exist in these models,
depending on the nature of the field potential \cite{Man2004,Shn2018,KevCue-Mar2019}. 

One very interesting phenomenon is a \emph{sphaleron} ---
a static, finite energy, unstable solution having a localized profile \cite{Man2019}. 
These solutions arise whenever there is
a saddle point between true and false vacua in the energy functional
\cite{AveBazLosMen}.
The field profile is symmetric with a single peak
given by the value of the true vacuum, 
and asymptotically the profile goes to a constant
which is equal to the value of the false vacuum. 

Sphalerons are known to exhibit a linear instability, 
which is easily understood to occur by the following argument \cite{Man2019} 
about the corresponding eigenfunction equation
that governs linear perturbations.
Consider the derivative of a sphaleron field,
which satisfies the linearized field equation
and thus represents an eigenfunction with zero eigenvalue, 
called a zero mode. 
This eigenfunction has a node at the location of the peak of the sphaleron. 
From a general result in eigenfunction theory (Sturm oscillation theorem),
the ground-state eigenfunction is always nodeless,
and therefore the ground state must have a negative eigenvalue.
Hence the sphaleron is linearly unstable when perturbed by the ground-state mode.

The evolution of perturbed sphalerons has been of substantial interest 
due to its relationship with long-lived localized oscillatory solutions 
called oscillons \cite{Man2019}. 
In particular, sphalerons have been observed numerically to collapse to oscillons,
and understanding this collapse process is important
for a clearer mathematical picture of oscillons
\cite{Bog.Mak,Gle,Gle.Sic,Nav-Obr.Que,Ada.Ciu.Ole.Rom.Wer.2021,Ole.Que.Rom.Wer.2023,Alo-Izq.Nav-Obr.Ole.Que.Rom.Wer.2023,Man.Rom.2023}.
To-date however,
a mathematical analysis of the instability modes of sphalerons
has yet to appear. 

The present paper is devoted to analytically studying the spectral problem for sphalerons. 
As main results, it shown that the eigenfunctions and eigenvalues 
can be obtained through transforming the problem into a Heun equation \cite{Ron-book},
which is a linear second-order differential equation that 
generalizes the hypergeometric equation.
In particular, the Heun equation has one additional regular singular point
compared to the hypergeometric equation. 
The eigenfunctions are found to be given by certain local Heun functions,
described by Frobenius series representations. 
Furthermore, the discrete spectrum is shown to comprise
the negative-eigenvalue ground state, and two positive-eigenvalue modes,
in addition to a standard zero mode arising from translation invariance.
As additional results,
an approximation in terms of elementary functions is found for
the ground state eigenvalue and eigenfunction.

Heun functions have attracted rising attention in the last two decades
as they have started to appear in a wide variety of advanced applications
in mathematical and theoretical physics \cite{Yu-Lay,Fiz.Sta,Hor},
including the topic of Schr\"odinger potentials \cite{Ish2018}
to which the spectral problem of sphalerons turns out to belong.

The rest of the present paper is organized as follows. 

In section~\ref{sec:sphaleron}, 
as a preliminary result,
it is shown that a general quartic Klein-Gordon potential with a false vacuum 
can be transformed into one-parameter family of potentials
by a shift, scaling, reflection on the Klein-Gordon field.
This family coincides with a potential studied in \Ref{AveBazLosMen}
and reviewed in \Ref{Man-review24}. 
The sphaleron solution is then introduced,
along with the the corresponding moving solitary wave solution
obtained by applying a Lorentz boost. 

In section~\ref{sec:stability}, 
the linear perturbation problem is studied for sphalerons. 
Its solution in terms of local Heun functions is presented,
using a change of variable motivated by
one of the singularities in the linear differential equation for the perturbation.
Both the eigenvalues and the eigenfunctions are derived analytically.
The discrete spectrum is also determined numerically in terms of local Heun functions. 

In section~\ref{sec:approx},
explicit approximate analytical expressions for the ground state eigenvalue and eigenfunction
are developed in terms of elementary functions.  

In section~\ref{sec:remarks},
some concluding remarks are given.

An appendix summarizes the conserved quantities
of nonlinear Klein-Gordon equations.

\section{Sphaleron in a false vacuum}
\label{sec:sphaleron}

For a general nonlinear KG model of a scalar field $\phi(x,t)$, 
the equation of motion is given in terms of an interaction potential $V(\phi)$ by 
\begin{equation}\label{KG.eqn}
\phi_{tt} - \phi_{xx} + V'(\phi) =0 , 
\end{equation}
which arises from the action principle
\begin{equation}\label{KG.action}
S[\phi] = \int_{-\infty}^\infty\int_{-\infty}^\infty\big( {-}\tfrac{1}{2}\phi_t^2 + \tfrac{1}{2}\phi_x^2 +V(\phi) \big)dx\,dt . 
\end{equation}
Here the field and the space-time coordinates are taken to be dimensionless 
(i.e.\ relativistic units are employed). 
There is freedom of adding an arbitrary constant $V_0$ to the interaction potential $V(\phi)$ 
without changing the equation of motion.
The action principle is invariant under time-translations, space-translations, and Lorentz boosts,
which yield respective conservation laws for energy, linear momentum, and boost momentum. 
See the appendix. 

Consider a non-symmetric quartic potential with a false vacuum. 
This means that $V(\phi)$ has a local maximum and two local minimums, 
one of which is a global minimum. 
In physical terms, the global minimum represents the true vacuum, 
while the other minimum represents the false vacuum. 
These conditions require, firstly, that $V'(\phi)$ must have three distinct real roots: 
$\phi = a_1, a_2, a_3$
where, without loss of generality, $a_1<a_2<a_3$. 
Secondly, $V'(\phi)$ must not be anti-symmetric when reflected around $\phi=a_2$:
$a_1 - a_2 \neq a_3 -a_2$. 
Thus, we must have 
$V'(\phi) = \alpha (\phi-a_1)(\phi-a_2)(\phi-a_3)$,
with $a_1<a_2<a_3$ and $a_2 \neq \tfrac{1}{2}(a_3 +a_1)$,
where $\alpha>0$. 

We now want to characterize this class of potentials 
by transforming $V'(\phi)$ into a simple canonical form. 
First, we can apply a shift transformation $\phi\to \phi -\tfrac{1}{2}(a_1+a_3)$ 
so that $V'(\phi) = \alpha (\phi^2 - \tilde b^2)(\phi -\tilde a)$, 
where 
$\tilde b = \tfrac{1}{2}(a_3-a_1) >0$
and
$\tilde a = a_2 - \tfrac{1}{2}(a_3 +a_1) \neq 0$,
which satisfy the inequality $\tilde b >|\tilde a|$. 
This yields 
\begin{equation}\label{V.general}
V(\phi) = \alpha\big( \tfrac{1}{4} \phi^4 -\tfrac{1}{3} \tilde a \phi^3 -\tfrac{1}{2} \tilde b^2 \phi^2 + \tilde a \tilde b^2 \phi \big) + V_0
\end{equation}
whose local extrema are given by 
$V(\tilde a) = V_0 + \tfrac{1}{12}\alpha (6 \tilde b^2 - \tilde a^2)\tilde a^2$ 
where $\phi=\tilde a$ is the maximum, 
and 
$V(\pm\tilde b) =V_0 -\tfrac{1}{12}\alpha (3\tilde b \mp 8\tilde a)\tilde b^3$
where $\phi=\pm \tilde b$ are the two minimums. 
The difference in the potential at these minima, 
which distinguishes the true vacuum from the false vacuum,  
is given by 
\begin{equation}
\Delta V := V(\tilde b) -V(-\tilde b)= \tfrac{4}{3}\alpha\tilde a \tilde b^3 . 
\end{equation}
There are two cases to consider, depending on sign of $\tilde a \neq0$. 

Case $\tilde a<0$: 
the false vacuum is $\phi = -\tilde b <0$ and the true vacuum is $\phi = \tilde b >0$,
where $-\tilde b< 0 < |\tilde a| < \tilde b$. 
As explained in the subsequent subsections, 
the shape of the potential \eqref{V.general} in this case 
indicates that it supports a solitary wave on a dark background, 
namely, $\phi \to -\tilde b$ at $x\to \pm\infty$. 
When this wave is stationary, it describes a ground state lump. 
By the following transformation, 
the background can be shifted to zero. 
We choose $V_0 = \tfrac{1}{12}\alpha\tilde b^3 (3\tilde b - 8|\tilde a|)$ 
and shift $\phi \to \phi +\tilde b$,
whereby  
\begin{equation}\label{V.general.neg_a}
  V(\phi) = \alpha \phi^2\big( \tfrac{1}{4} \phi^2 +(\tfrac{1}{3} |\tilde a| -\tilde b)\phi + \tilde b(\tilde b -|\tilde a|) \big),
  \quad
  \tilde b > -\tilde a >0
\end{equation}
with $V'(\phi) = \alpha \phi (\phi - 2\tilde b)(\phi -\tilde b +|\tilde a|)$. 
In this shifted potential, the false vacuum is $\phi=0$, 
and the true vacuum is $\phi = 2\tilde b>0$. 
The ground state lump is bright,
with $\phi$ going from $0$ to a peak
$\phi_{\max} = \tfrac{2}{3}( 3\tilde b -|\tilde a| - \sqrt{|\tilde a|(3b + |\tilde a|)}\, ) >0$,
which is the smallest positive root of the potential. 

Case $\tilde a>0$: 
the vacua are interchanged,
with $\phi =\tilde b$ being the false vacuum
and $\phi = -\tilde b$ the true vacuum, 
where $-\tilde b< 0<\tilde a < \tilde b$. 
The potential \eqref{V.general} in this case supports
a solitary wave on a bright background, $\phi \to \tilde b$ at $x\to \pm\infty$. 
This wave describes a ground state lump when it is stationary. 
Using similar steps as in the previous case, 
we can shift the potential to get 
\begin{equation}\label{V.general.pos_a}
  V(\phi) = \alpha \phi^2\big(\tfrac{1}{4} \phi^2 +(\tilde b - \tfrac{1}{3} \tilde a) \phi + \tilde b(\tilde b-\tilde a) \big),
  \quad
  \tilde b > \tilde a >0
\end{equation}
with $V'(\phi) = \alpha \phi (\phi +2\tilde b)(\phi +\tilde b -\tilde a)$
so that the false vacuum is again $\phi =0$ 
while the true vacuum is $\phi = -2\tilde b<0$. 
The ground state lump is now dark,
with $\phi$ going from
$\phi_{\min} = -\tfrac{2}{3}( 3\tilde b -\tilde a -\sqrt{\tilde a(3b + \tilde a)}\, ) <0$ to $0$.

It will now be useful to express the potentials \eqref{V.general.neg_a} and \eqref{V.general.pos_a}
in a simpler factorized form by a reparameterization.
This also leads to a very simple expression for the ground state lumps,
known as sphalerons. 
Hereafter we will focus on the potential \eqref{V.general.neg_a} 
in which the ground state lump is bright. 
The opposite case in which the ground state lump is dark 
is given by a reflection $\phi \to -\phi$, corresponding to the potential \eqref{V.general.pos_a}. 

The reparameterization of the potential \eqref{V.general.neg_a} 
is achieved by expressing its two positive roots as 
\begin{equation}
\tfrac{2}{3}\big( 3\tilde b - |\tilde a| -\sqrt{|\tilde a|(3\tilde b + |\tilde a|)} \big)
= b\,\tanh(a), 
\quad
\tfrac{2}{3}\big( 3\tilde b - |\tilde a| +\sqrt{|\tilde a|(3\tilde b + |\tilde a|)} \big)
= b\,\coth(a) 
\end{equation}
where $a>0$ and $b>0$. 
Equivalently, we have 
\begin{equation}
b = 2\sqrt{\tilde b(\tilde b -|\tilde a|)},
\quad
a = \tfrac{1}{2}\arctanh\Big(\sqrt{1-|\tilde a|/\tilde b}\big/\big(1-\tfrac{1}{3}|\tilde a|/\tilde b\big)\Big) . 
\end{equation}
A scaling transformation $\phi \to b\phi$ combined with 
a dilation $(x,t) \to \sqrt{\tfrac{2}{\alpha}}(x,t)$
then leads to the following characterization of the resulting potential. 

\begin{prop}\label{prop:potential.falsevacuum}
Up to a shift, scaling, and reflection on $\phi$, and a dilation on $(t,x)$, 
any non-symmetric quartic potential with a false vacuum 
belongs to the $1$-parameter family 
\begin{equation}\label{potential}
V(\phi) = 2 \phi^2 (\phi - \tanh(a))(\phi -\coth(a)), 
\quad
a>0 . 
\end{equation}
The false vacuum is $V=0$ at $\phi=0$,
and the true vacuum has $V<0$ at 
\begin{equation}\label{true.vacuum}
\phi = \tfrac{3}{4} \coth(2a) \Big( 1 + \sqrt{1 - \tfrac{8}{9}\tanh(2a)^2} \Big)
\end{equation}
which lies between $\phi=\tanh(a)$ and $\phi=\coth(a)$. 
The equation of motion \eqref{KG.eqn} in this potential is given by 
\begin{equation}\label{eom.potential}
\phi_{tt} - \phi_{xx} + 4(2\phi^2 -3\coth(2a)\phi + 1)\phi =0 . 
\end{equation}
\end{prop}

\begin{figure}
\includegraphics[width=0.5\textwidth,trim=2cm 12cm 2cm 1cm, clip]{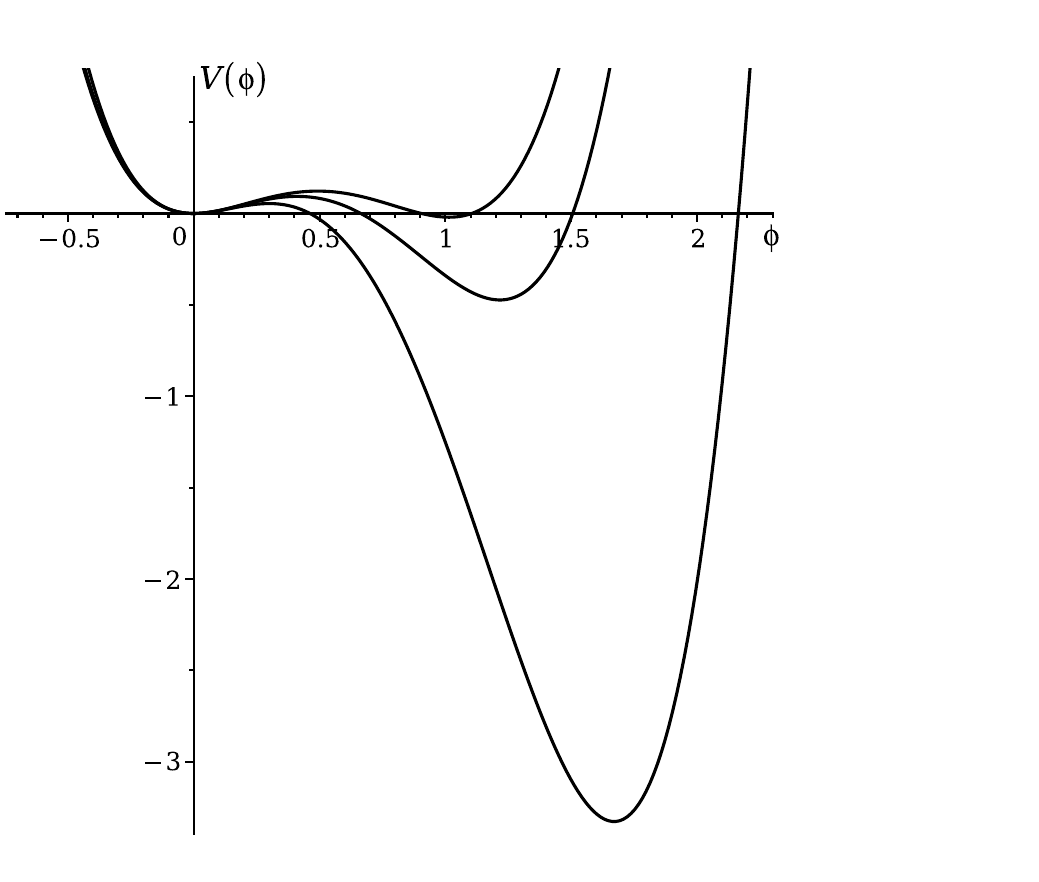}
\caption{Potential \eqref{potential} with a false vacuum at $\phi=0$: 
$a=$ 0.5, 0.8, 1.5}
\label{fig:potential}
\end{figure}

This family \eqref{potential} is equivalent, up to scaling, to the potential 
\begin{equation}\label{potential1}
V(\phi) = 2\phi^2 (\phi -\tilde a)(\phi-\tilde a/\tilde b)
\end{equation}
with $1>\tilde b>0$, $\tilde a>0$ studied in \Ref{AveBazLosMen},
and also to the potential 
\begin{equation}\label{potential2}
V(\phi) = 4 \phi^2 (1 -\phi/\tilde b^2)(1-\phi/\tilde a^2)
\end{equation}
with $\tilde b >\tilde a >0$ listed in \Ref{Man-review24}.
A plot of family \eqref{potential} is shown in Fig.~\ref{fig:potential}. 

\subsection{Ground states}\label{sec:groundstate}

A ground state of a general nonlinear KG model \eqref{KG.eqn}
is a static solution $\phi=\phi(x)$ 
satisfying the ODE 
\begin{equation}\label{groundstate.eqn}
\phi_{xx} = V'(\phi) . 
\end{equation}
This is an oscillator equation with effective potential $V_\eff(\phi)= -V(\phi)$. 
Integration gives 
\begin{equation}\label{groundstate.energy.eqn}
\tfrac{1}{2}\phi_x^2  - V(\phi) =E_0
\end{equation}
where $E_0=\const$ is the oscillator energy. 

The shape of the effective potential 
determines the allowed non-constant ground states $\phi(x)$. 
For a non-symmetric potential \eqref{potential}, 
the effective oscillator potential is an inverted quartic 
\begin{equation}\label{oscil.potential}
V_\eff(\phi) = - 2\phi^2 (\phi -\tanh(a))(\phi -\coth(a))
\end{equation}
which has 
a local maximum at $\phi=0$ 
and a global maximum at 
\begin{equation}
\phi = \tfrac{3}{4} \coth(2a) \big( 1 + \sqrt{1 - \tfrac{8}{9}\tanh(2a)^2} \big), 
\end{equation}
in addition to a local minimum at 
\begin{equation}
\phi = \tfrac{3}{4} \coth(2a) \big( 1 - \sqrt{1 - \tfrac{8}{9}\tanh(2a)^2} \big) >0 . 
\end{equation}

It is straightforward to see that 
the only non-constant ground states supported in this potential \eqref{oscil.potential}
consist of a bright sphaleron 
where $\phi$ goes from $0$ at $x=\pm\infty$ to a peak $\phi_{\max} = \tanh(a)$ 
at some position $x=x_0$. 
This sphaleron thereby has $E_0=0$. 
Integration of the oscillator equation \eqref{groundstate.energy.eqn}
yields the explicit expression 
\begin{equation}\label{lump}
\phi(x) = \sinh(2a)/\big( \cosh(2a) + \cosh(2(x-x_0)) \big) . 
\end{equation}

\begin{figure}
\includegraphics[width=0.75\textwidth,trim=2cm 12cm 2cm 1cm, clip]{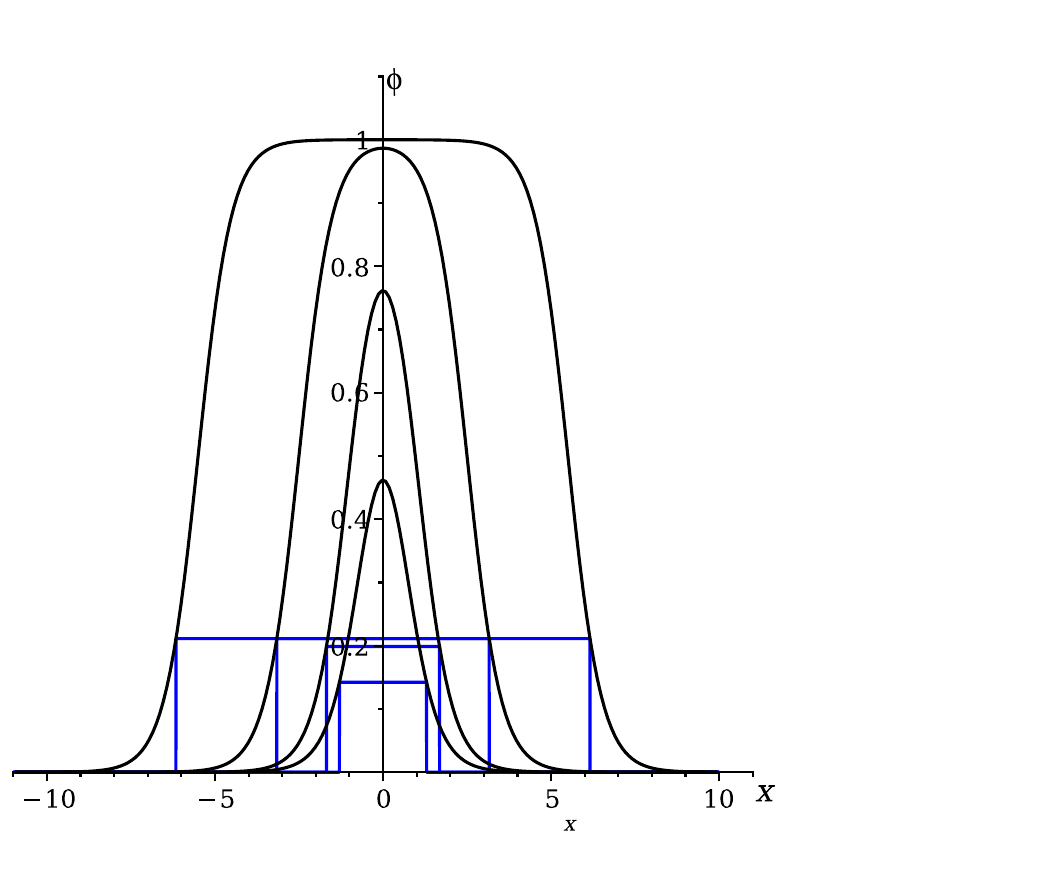}
\caption{Lump solution \eqref{lump} in the false-vacuum potential \eqref{potential}
for $a= $ 5.5, 2.5, 1.0, 0.5}
\label{fig:lump}
\end{figure}

This solution was first derived in a slightly different form 
for the equivalent potentials \eqref{potential1} and \eqref{potential2}. 
In particular, 
$\phi(x)=\tilde b/\big(1+(1-\tilde a)\sinh(\tilde b x/\sqrt{\tilde a})^2\big)$
and 
$\phi(x) = \tilde a^2\tilde b^2/\big(\tilde a^2 +(\tilde b^2 -\tilde a^2)\cosh(x)^2\big)$
appear respectively in \Ref{AveBazLosMen} and \Ref{Man-review24}. 
The latter also states another very useful form of the solution,
which is given by 
\begin{equation}\label{kinkantikink}
\phi(x) = \tfrac{1}{2} \big(\tanh(x-x_0+a) - \tanh(x-x_0 -a)\big) 
\end{equation}
for the potential \eqref{potential}. 
This shows that the sphaleron \eqref{lump} can be viewed as 
the superposition of a kink and an antikink.
It has a peak height
\begin{equation}\label{height}
\phi(x_0)= \tanh(a)
\end{equation}
and its full-width is 
\begin{equation}\label{width}
\Delta x = \arccosh\big(2\cosh(2a) + \sqrt{3\cosh(2a)^2 + 6}\big)
\end{equation}
as defined by where the convexity of $\phi(x)$ is maximum.
See Fig.~\ref{fig:lump}. 

The energy of a ground state is given by the integral 
\begin{equation}
E[\phi] = \int_{-\infty}^{\infty} \Big( \tfrac{1}{2} \phi_x^2 + V(\phi) \Big)\, dx 
= 2 \int_{-\infty}^{\infty} V(\phi) \, dx . 
\end{equation}
Since the sphaleron \eqref{lump} has a single peak and decays to zero,
its energy can be expressed as a Bogomolny integral 
\begin{equation}
E[\phi] = 2\int_{0}^{\phi_{\max}} \sqrt{2V(\phi)} \, d\phi
\end{equation}
where $\phi_{\max} = \tanh(a)$.
This yields 
\begin{equation}\label{lump.E}
E_\lump = \tfrac{2}{3} + 2\big(1 - 2a\coth(2a)\big)/\sinh(2a)^2
\end{equation}
for the energy. 
(A more complicated expression for the energy is given in \Ref{Man-review24}
using the alternative form \eqref{potential2} for the potential.)
Because the lump is stationary, it clearly has zero momentum, $P_\lump=0$, 
and zero boost-momentum, $J_\lump=0$.

\subsection{Travelling waves}\label{sec:travel.wave}

A Lorentz boost can be applied to any static solution 
$\phi=\phi(x)$ of the equation of motion, 
yielding a travelling wave solution $\phi=\phi((x -\nu t)/\sqrt{1-\nu^2})$
with speed $\nu$. 
Travelling waves move rightward if $\nu >0$, or leftward if $\nu<0$. 
A physical requirement is that $|\nu|<1$, 
so their motion lies inside the light cone; solutions with $|\nu|>1$ are called tachyonic. 

The sphaleron \eqref{lump} gives rise to solitary waves
\begin{equation}\label{solitarywave}
\phi(x,t) = \sinh(2a)/\big( \cosh(2a) + \cosh(2(\xi -\xi_0)) \big)
\end{equation}
in terms of travelling wave variable
\begin{equation}
\xi = \frac{x -\nu t}{\sqrt{1-\nu^2}}, 
\quad
|\nu| < 1 . 
\end{equation}
An equivalent expression for the solitary waves is given by 
\begin{equation}\label{kinkantikinkwave}
\phi(x,t) = \tfrac{1}{2} \big(\tanh(\xi-\xi_0 +a) - \tanh(\xi-\xi_0 -a)\big) . 
\end{equation}
This describes the superposition of a kink and an antikink which move in the same direction. 

The solitary wave has energy 
\begin{equation}\label{solitary.E}
E_\solitary = \frac{E_\lump}{\sqrt{1-\nu^2}}
= \frac{2}{\sqrt{1-\nu^2}}\Big( \tfrac{1}{3} + \big(1 - 2a\coth(2a)\big)/\sinh(2a)^2 \Big)
\end{equation}
and momentum 
\begin{equation}\label{solitary.P}
P_\solitary = \frac{\nu E_\lump}{\sqrt{1-\nu^2}}
= \frac{2\nu}{\sqrt{1-\nu^2}}\Big( \tfrac{1}{3} + \big(1 - 2a\coth(2a)\big)/\sinh(2a)^2 \Big) . 
\end{equation}
These expressions give an energy-momentum 2-vector $(E_\solitary,P_\solitary)$ 
which is the boost of the 2-vector $(E_\lump,0)$.

\section{Linear Instability}
\label{sec:stability}

To investigate linear stability of a ground state $\phi=\phi(x)$, 
consider a small perturbation with frequency $\omega$, 
\begin{equation}\label{perturbed}
\varphi(x,t)=\phi(x)+ \epsilon \eta(x)\exp(i\omega t)
\end{equation}
where $|\epsilon|\ll 1$. 
Linearization of the equation of motion 
for $\varphi(x,t)$ around $\epsilon=0$ yields
\begin{equation}
0 = \varphi_{tt} - \varphi_{xx} + V'(\varphi) 
= \epsilon \exp(i\omega t) \Big( {-}\omega^2 \eta - \eta_{xx} + V''(\phi(x)) \eta)\Big)
+ O(\epsilon^2) , 
\end{equation}
which gives a linear Schr\"{o}dinger equation for $\eta(x)$: 
\begin{equation}\label{perturbation.eqn}
-\eta_{xx} +U(x)\eta =\omega^2\eta
\end{equation}
with the potential
\begin{equation}\label{perturbation.potential}
U(x)= V''(\phi(x)) . 
\end{equation}
This is an eigenfunction problem for the linear operator $-\partial_x^2 + U(x)$, 
with eigenvalue $\lambda = \omega^2$. 

If all eigenvalues are non-negative, then $\omega$ is real 
and thus the perturbation \eqref{perturbed} is oscillatory in $t$, 
whereby the ground state $\phi(x)$ would be linearly stable. 
A zero eigenvalue, $\lambda_0=0$, is given by the eigenfunction $\eta(x)= \phi'(x)$,
which is called the translational mode or zero mode. 
If at least one negative eigenvalue exists, $\lambda_{-1} <0$, 
it gives $\omega = \pm i\sqrt{|\lambda_{-1}|}$,
which implies the perturbation \eqref{perturbed} is exponential in time,
since $\exp(i\omega t) = \exp(\pm \sqrt{|\lambda_{-1}|}\,t)$. 
Existence of the growing mode $\eta(x)\exp(\sqrt{|\lambda_{-1}|})$  
would indicate that the ground state $\phi(x)$ is linearly unstable. 

The goal is now to prove rigorously that the sphaleron \eqref{lump}
in the false vacuum of the potential \eqref{potential} 
is linearly unstable.

\subsection{Spectrum of the perturbation potential}\label{sec:perturbation.potential}

\begin{figure}
\includegraphics[width=0.75\textwidth,trim=2cm 12cm 2cm 1cm, clip]{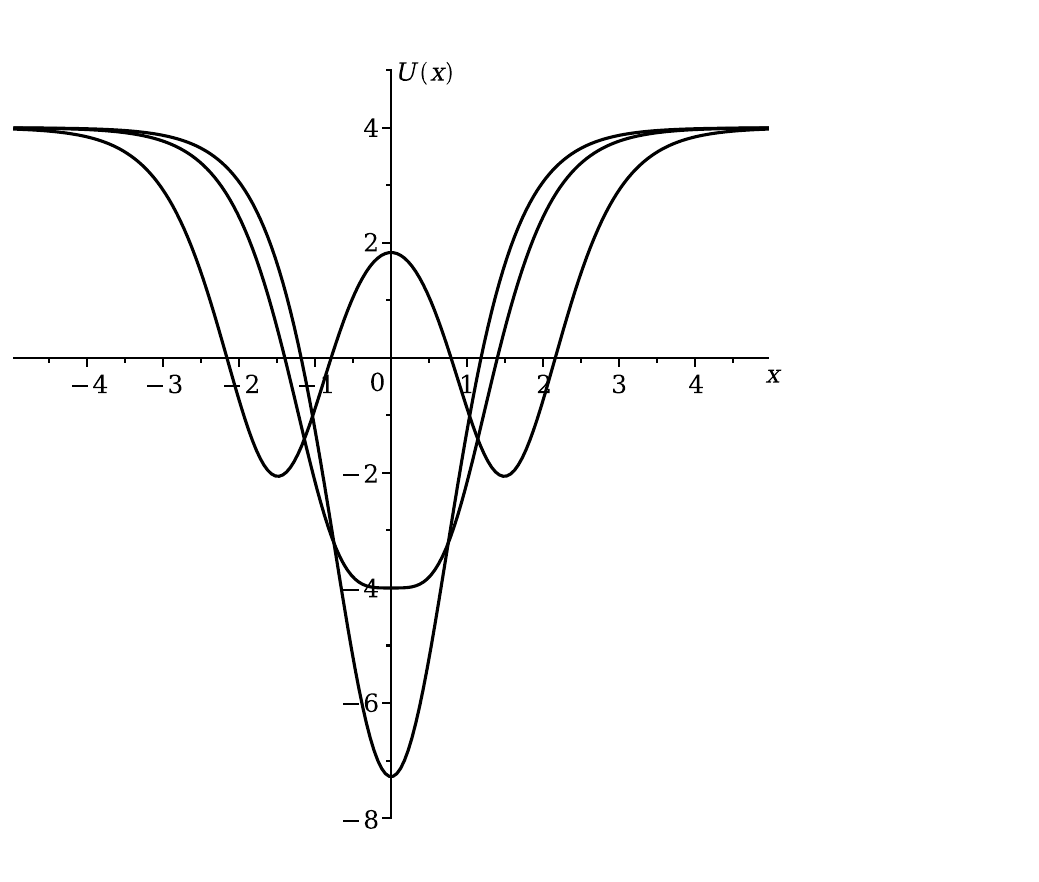}
\caption{Perturbation potential: 
$a=$ 0.25, $\arctanh\tfrac{1}{\sqrt{3}}$ (=0.658), 1.5}
\label{fig:U.well}
\end{figure}

The potential \eqref{potential} has 
$V''(\phi(x)) = 4(6\phi(x)^2 -3(\tanh(a) + \coth(a))\phi(x) + 1)$. 
For the sphaleron \eqref{lump}, 
this yields the perturbation potential 
\begin{equation}\label{U}
U(x) = 4 - 24\frac{1 +\cosh(2x)\cosh(2a)}{(\cosh(2x)+\cosh(2a))^2} 
\end{equation}
which has the following main features. 
It is asymptotically constant, with $U(x)\to 4$ exponentially as $|x|\to\infty$, 
while $U(x) <4$ for finite $x$. 
It is symmetric in $x$, such that $x=0$ is the minimum when $\coth(a) \geq \sqrt{3}$ 
or the local maximum when $\coth(a) < \sqrt{3}$,
thus giving a double-well in the latter case and a single-well in the former case. 
See Fig.~\ref{fig:U.well}. 
In addition, the  zero mode of this potential is given by
\begin{equation}\label{zeromode} 
\eta(x)= \phi'(x) = -2\sinh(2a) \sinh(2x)/(\cosh(2x) + \cosh(2a))^2,
\quad
\lambda_0 =0
\end{equation}
which has a single node, $\eta=0$ at $x=0$.

These features allow the spectrum of $-\partial_x^2 + U(x)$
to be deduced from basic results about linear Schr\"odinger operators (see e.g.\ \Ref{GeyPel-book}). 
Firstly, the continuous spectrum consists of $[4,\infty)$, by Weyl's theorem. 
Secondly, the lowest eigenvalue in the discrete spectrum will be negative,
$\lambda_{-1}<0$, 
corresponding to a nodeless eigenfunction $\eta_{-1}(x)$,
due to Sturm's oscillation theorem. 

Therefore, the sphaleron \eqref{lump} is linearly unstable. 
Remarkably, the eigenfunction problem 
\begin{equation}\label{lump.perturbation.eqn}
-\eta_{xx} +U(x)\eta =\lambda\eta,
\quad
\lambda = \omega^2
\end{equation}
can be solved in terms of Heun functions,
as will be shown next through an explicit transformation. 
This will lead to an analytical way to find the negative eigenvalue $\lambda_{-1}$. 

To begin,
we will examine the singular points of the linear differential equation \eqref{lump.perturbation.eqn}.
The potential \eqref{U} is singular at the finite points
$x_0= \pm a + i \pi/2$ (modulo $\pi$),
and each of them is a regular singular point,
since $(x-x_0)^2 U(x)$ is analytic around $x=x_0$. 
Furthermore, the point at infinity, $x=\infty$, is irregular singular, 
since the potential is not analytic in $1/x$. 
We can transform this point into a regular singular point
by a change of variable of the form $x\to y = f(\cosh(x))$ where $f$ is some rational function. 

The most effective choice turns out to be given by the change of variable 
\begin{equation}\label{chgvars.y}
y=1/\cosh^2(x)
\end{equation}
under which $U(x)$ becomes the following rational function of $y$: 
\begin{equation}
U = 4 + \frac{12 y(\sinh^2(a) y - \cosh(2a))}{(\sinh^2(a) y +1)^2} . 
\end{equation}
Since $\dfrac{dy}{dx} = -2 y \sqrt{1-y}$, 
the eigenfunction equation \eqref{lump.perturbation.eqn} 
is correspondingly transformed to the rational form 
\begin{equation}\label{lump.eigenfunct.eqn.y}
\tilde\eta''(y) + \frac{3y -2}{2y(y - 1)} \tilde\eta'(y) 
+ \Big( 
\frac{4-\lambda}{4y^2(y -1)}
+ \frac{3(\sinh^2(a) y -\cosh(2a))}{y(y - 1)(\sinh^2(a) y + 1)^2}
\Big) \tilde\eta(y) 
=0
\end{equation}
for the function 
\begin{equation}\label{lump.eigenfunct.y}
\tilde\eta(y) = \eta(x)= \eta(\arccosh(1/\sqrt{y})) . 
\end{equation}
We see that coefficients of $\tilde\eta'(y)$ and $\tilde\eta(y)$ 
are singular at $y=0,1,-1/\sinh^2(a)$ and also at $y=\infty$. 
These are readily checked to be regular singular points of the linear ODE \eqref{lump.eigenfunct.eqn.y}.

Now recall that any second-order linear ODE with four regular singular points 
is equivalent under a change of variables to Heun's ODE \cite{Ron-book,Ish2018}
\begin{equation}\label{heun.ode}
H''(z) 
+  \big( \gamma/z + \delta/(z-1) +\epsilon/(z-p) \big)H'(z) 
+\big( (\alpha\beta z -q)/(z(z-1)(z-p) \big) H(z)
=0
\end{equation}
with $\epsilon = \alpha + \beta -\gamma -\delta + 1$,
which has regular singular points $z=0,1,p,\infty$, 
where $\alpha$, $\beta$, $\gamma$, $\delta$, and $q$ are constant parameters.
Solutions $H(z)$ are given by an analytic Frobenius series around any one of the points $z=0,1,p$,
in terms of the local Heun function $H\ell(p,q,\alpha,\beta,\gamma,\delta;z)$
which is analytic in $z$ with $H\ell(p,q,\alpha,\beta,\gamma,\delta;0)=1$.
The main properties of local Heun functions are summarized in \Ref{Ron-book}.

The explicit transformation of the linear ODE \eqref{lump.eigenfunct.eqn.y} to Heun's ODE \eqref{heun.ode} 
is given by 
\begin{equation}\label{heun.y}
H(y) = \frac{\tilde\eta(y)}{y^{\sqrt{1-\lambda/4}}(\sinh(a)^2 y + 1)^3} 
\end{equation}
with 
\begin{equation}\label{Heun.ode.params}
\alpha = \tfrac{1}{2}\sqrt{4-\lambda} + 3,
\quad
\beta = \tfrac{1}{2}(\sqrt{4-\lambda} + 7),
\quad
\gamma = \sqrt{4-\lambda} + 1,
\quad
\delta =\tfrac{1}{2},
\quad
\epsilon = 6
\end{equation}
and
\begin{equation}\label{Heun.ode.p.q}
p = -\frac{1}{\sinh^2(a)},
\quad
q = \frac{6\cosh(2a)(\sqrt{4-\lambda} +3) -(\sqrt{4-\lambda} +6)(\sqrt{4-\lambda} +1)}{4\sinh^2(a)} . 
\end{equation}
In terms of $x$, 
we have
\begin{equation}
\eta(x) = \sech(x)^{\sqrt{4-\lambda}}(\sinh(a)^2 \sech^2(x) + 1)^3 H(\sech^2(x)) , 
\end{equation}
where the finite singular points $y=0,1,-1/\sinh^2(a)$ 
respectively correspond to $x=\pm\infty,0,\pm a + i\pi/2$ (modulo $\pi$). 

From the general features of the potential \eqref{U}, we can deduce that 
the eigenfunction $\eta_{-1}(x)$ with the negative eigenvalue $\lambda_{-1}$
must be analytic in $x$, and decay like $e^{-\sqrt{4 +|\lambda_{-1}|}\,|x|}$ as $|x|\to\infty$, 
in addition to being symmetric around $x=0$ and nodeless. 
In particular, as a consequence of analyticity,
the symmetry $\eta_{-1}(-x) = \eta_{-1}(x)$ will hold if and only if $\eta_{-1}'(0)=0$. 
Through expression \eqref{heun.y} 
combined with the transformation \eqref{chgvars.y},
the preceding properties imply that \\
(i) 
$H(y)$ must be analytic in $y$, 
non-singular at $y=0$, non-zero for $0<y<1$;
\\
(ii) 
$\sqrt{1-y} \big( y^{\sqrt{1-\lambda/4}}(\sinh(a)^2 y + 1)^3 H(y) \big)'$ must vanish at $y=1$. 

Condition (i) implies that, after normalization $H(0)=1$, 
the function $H(y)$ must be equal to the local Heun function:  
\begin{equation}\label{H(y)}
H(y) = H\ell(p,q;\alpha,\beta,\gamma,\delta;y) . 
\end{equation}
Around $y=1$, this function \eqref{H(y)} has a Frobenius series expansion in $1-y$ 
with indicial exponents $0$ and $1-\delta = 1/2$ 
(using parameters \eqref{Heun.ode.params}):
\begin{equation}\label{H(y).alt}
\begin{aligned}
H(y) = & 
C_1 H\ell(1-p,\alpha\beta - q;\alpha,\beta,\delta,\gamma;1-y) 
\\&
+ C_2 \sqrt{1-y} H\ell(1-p,\alpha'\beta' - q';\alpha',\beta',\delta',\gamma;1-y) 
\end{aligned}
\end{equation}
where
\begin{equation}
q' = q +p\gamma (1-\delta), 
\quad
\alpha' = \alpha-\delta+1, 
\quad
\beta' = \beta-\delta+1,
\quad
\delta' = 2-\delta . 
\end{equation}
Thus, $H(1) = H\ell(p,q;\alpha,\beta,\gamma,\delta;1)$ is a finite value,
due to 
$H\ell(1-p,\alpha\beta - q;\alpha,\beta,\delta,\gamma;0)
= H\ell(1-p,\alpha'\beta' - q';\alpha',\beta',\delta',\gamma;0) 
=1$. 
Condition (ii) thereby yields, after simplification, 
\begin{equation}\label{eigenfunct.condition}
\lim_{y\to 1} \sqrt{1-y}H\ell(p,q;\alpha,\beta,\gamma,\delta;y)' = 0
\end{equation}
which constitutes the determining equation 
for the negative eigenvalue $\lambda=\lambda_{-1}$. 
This equation can be expressed in a useful equivalent form as follows. 
First,
via substituting the relation \eqref{H(y).alt} and using the finiteness of $H(1)$, 
we obtain $C_2 =0$. 
Next, we evaluate relation \eqref{H(y).alt} for $y=0$ and $y=1$, 
which gives 
\begin{equation}\label{C1.C2}
\begin{aligned}
C_1 & = H\ell(p,q;\alpha,\beta,\gamma,\delta;1)
\\
C_2 & = \big( 1 - C_1 H\ell(1-p,\alpha\beta - q;\alpha,\beta,\delta,\gamma;1) \big)/H\ell(1-p,\alpha'\beta' - q';\alpha,'\beta',\delta',\gamma;1)
\end{aligned}
\end{equation}
where
\begin{equation}
\begin{gathered}
q' = \frac{6\cosh(2a)(\sqrt{4-\lambda} +3) -(\sqrt{4-\lambda} +8)(\sqrt{4-\lambda} +1)}{4\sinh^2(a)},
\\
\alpha' = \tfrac{1}{2}(\sqrt{4-\lambda} + 7),
\quad
\beta' = \tfrac{1}{2}\sqrt{4-\lambda} + 4,
\quad
\delta' = \tfrac{3}{2} . 
\end{gathered}
\end{equation}
Finally, the denominator in expression \eqref{C1.C2} for $C_2$ 
turns out to be finite and non-zero. 
Hence, the determining equation $C_2=0$ becomes 
\begin{equation}\label{eigenvalue.deteqn}
H\ell(p,q;\alpha,\beta,\gamma,\delta;1) H\ell(1-p,\alpha\beta - q;\alpha,\beta,\delta,\gamma;1) 
-1 =0
\end{equation}
in terms of the parameters \eqref{Heun.ode.params}.
This establishes the first main result. 

\begin{figure}
\includegraphics[width=0.75\textwidth,trim=2cm 12cm 2cm 1cm, clip]{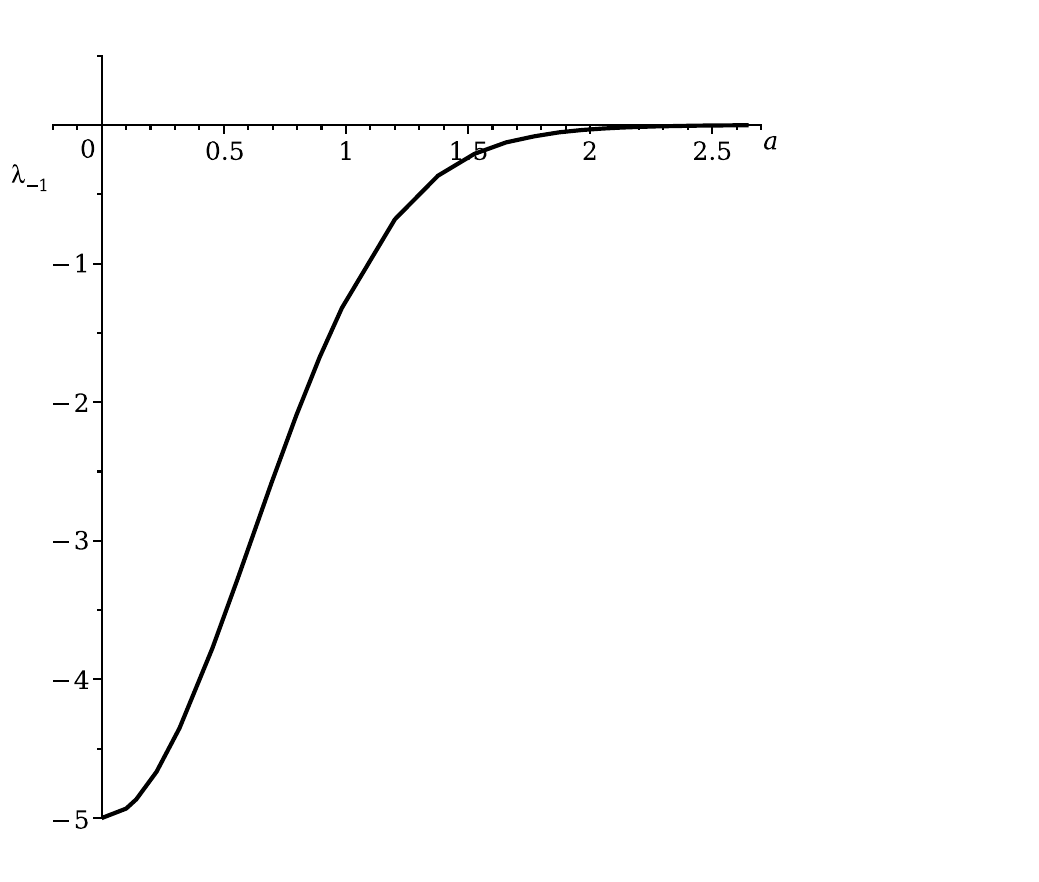}
\caption{Ground state eigenvalue: $\lambda_{-1}$}
\label{fig:neg.eigenvalue}
\end{figure}

\begin{thm}
The ground state eigenvalue $\lambda_{-1}$ of the potential \eqref{U}
satisfies equation \eqref{eigenvalue.deteqn}
in terms of the local Heun function $H\ell(p,q,\alpha,\beta,\gamma,\delta;z)$. 
For any fixed value of the parameter $a$ in the potential, 
this equation \eqref{eigenvalue.deteqn}
can be solved numerically. 
A plot of $\lambda_{-1}$ as a function of $a$ is shown in Fig.~\ref{fig:neg.eigenvalue}.
\end{thm}

From a numerical plot,
we see that there is a single negative solution $\lambda_{-1}$
for each value of the parameter $a$. 
The resulting solution of the eigenfunction equation \eqref{lump.eigenfunct.eqn.y} 
is given by 
\begin{equation}\label{-1.eigenfunct.y}
\tilde\eta_{-1}(y) = 
y^{\sqrt{1-\lambda_{-1}/4}} (\sinh(a)^2 y + 1)^3 H\ell(p,q;\alpha,\beta,\gamma,\delta;y) \big|_{\lambda=\lambda_{-1}} . 
\end{equation}

\begin{figure}
\includegraphics[width=0.8\textwidth,trim=2cm 12cm 2cm 1cm, clip]{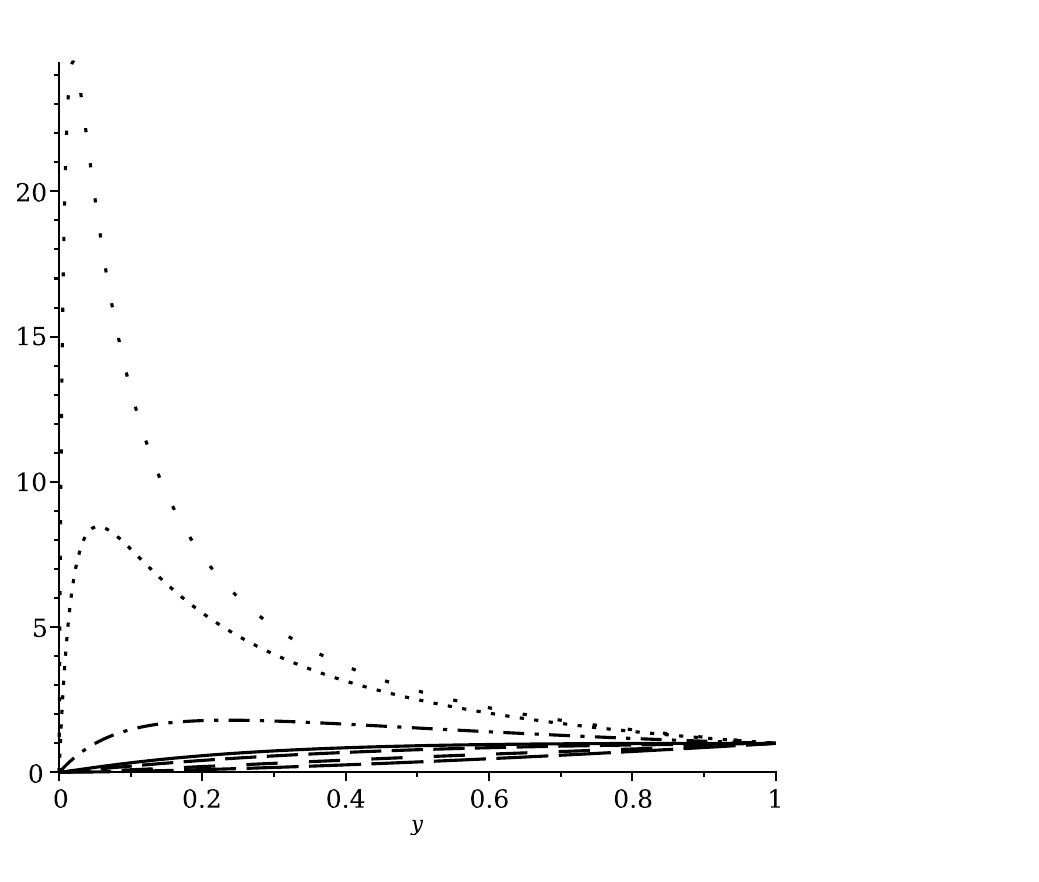}
\caption{$\tilde\eta_{-1}(y)$ 
for $a=$ 2.65 (space dot), 2.14 (dot), 1.44 (dash dot), 1.00 (solid), 0.881 (dash), 0.647 (space dash), 0.100 (long dash)}
\label{fig:neg.eigenvalue.H(y)}
\end{figure}

For numerical purposes, 
we note that the power series in $y$ which defines the Heun function \eqref{H(y)} 
has radius of convergence $\min(1,|p|)$,
where $|p|=1/\sinh(a)^2 <1$ when $a> \arctanh(\tfrac{1}{\sqrt{2}})$.
Consequently, for an arbitrary value of $a$, 
the series does not cover the whole interval $0< y\leq 1$. 
However, from the relation \eqref{H(y).alt} with $C_2=0$, 
we see that the Heun function 
\begin{equation}\label{H(y).numeric}
H(y) = H\ell(1-p,\alpha\beta - q;\alpha,\beta,\delta,\gamma;1-y) 
\end{equation}
differs only by the factor $C_1$ 
while the radius of convergence of its defining power series in $1-y$ 
is $\min(1,|1-p|) =1$ since $1-p = 1/\tanh(a)^2\geq 1$. 
Thus, for any value of $a$, the whole interval $0<y\leq 1$ is covered.
Using this function, we obtain the equivalent (up to normalization)
eigenfunction 
\begin{equation}\label{-1.eigenfunct.y.imprv}
\tilde\eta_{-1}(y) = 
y^{\sqrt{1-\lambda_{-1}/4}} (\sinh(a)^2 y + 1)^3 H\ell(1-p,\alpha\beta - q;\alpha,\beta,\delta,\gamma;1-y)  \big|_{\lambda=\lambda_{-1}} . 
\end{equation}
A plot is shown Fig.~\ref{fig:neg.eigenvalue.H(y)}. 

The corresponding eigenfunction is given by expressions \eqref{lump.eigenfunct.y}, \eqref{heun.y}, \eqref{H(y).numeric},
which leads to the following main result. 

\begin{figure}
\includegraphics[width=0.75\textwidth,trim=2cm 12cm 2cm 1cm, clip]{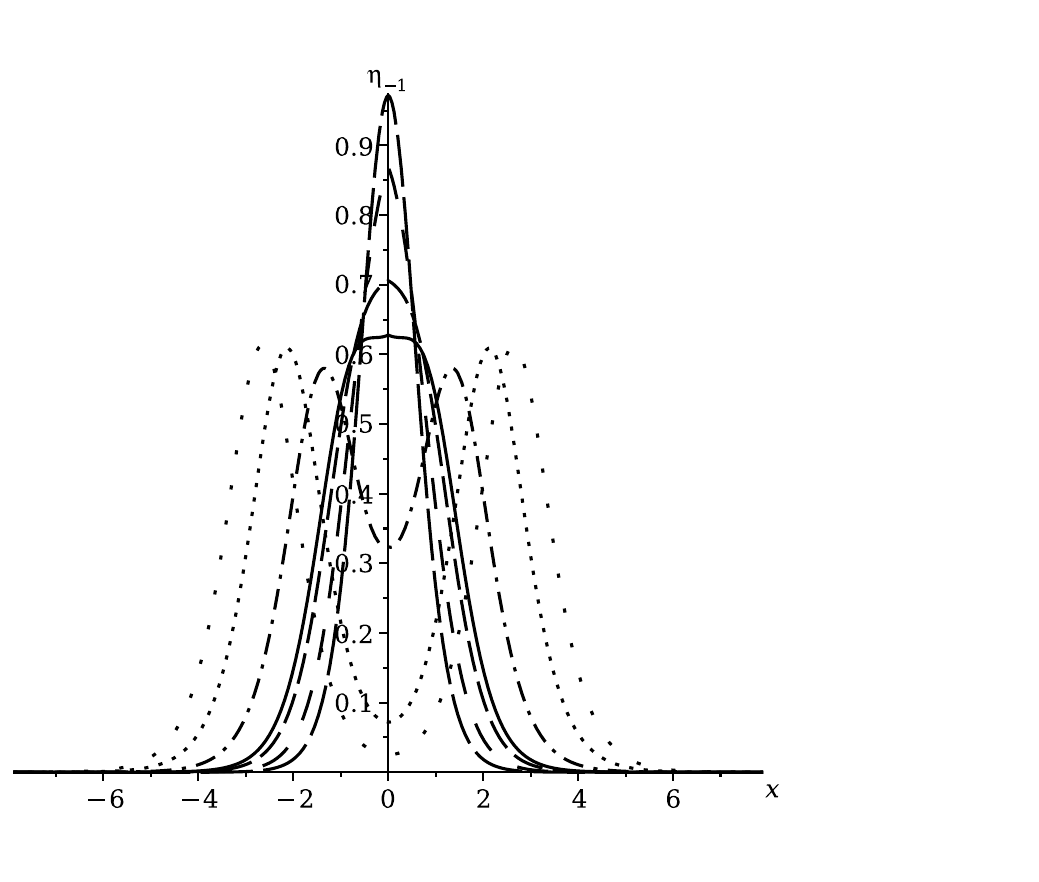}
\caption{Ground-state eigenfunction, normalized in $L^2$, 
for $a=$ 2.65 (space dot), 2.14 (dot), 1.44 (dash dot), 1.00 (solid), 0.881 (dash), 0.647 (space dash), 0.100 (long dash)}
\label{fig:neg.eigenfunction.eta.x}
\end{figure}

\begin{thm}
The ground state eigenfunction has the form 
\begin{equation}\label{-1.eigenfunct.x}
\eta_{-1}(x) = 
\sech(x)^{\sqrt{4-\lambda_{-1}}} (\sinh(a)^2 \sech^2(x) + 1)^3 
H\ell(1-p,\alpha\beta - q;\alpha,\beta,\delta,\gamma;\tanh^2(x))\big|_{\lambda=\lambda_{-1}}
\end{equation}
in terms of the parameters \eqref{Heun.ode.params}
and the local Heun function $H\ell$. 
A numerical plot of the normalized ground state is shown in Fig.~\ref{fig:neg.eigenfunction.eta.x}
for various values of the parameter $a$ in the potential \eqref{U}. 
\end{thm}

These eigenfunctions have a single peak, at $x=0$, for $a\lesssim 1.00$, and a double peak for $a\gtrsim 1.00$.
For $a\simeq 1.00$, when the eigenfunction has no convexity at $x=0$,
the corresponding eigenvalue is $\lambda_{-1}\simeq 1.30$. 

\begin{figure}
\includegraphics[width=0.46\textwidth,trim=2cm 12cm 4cm 1cm, clip]{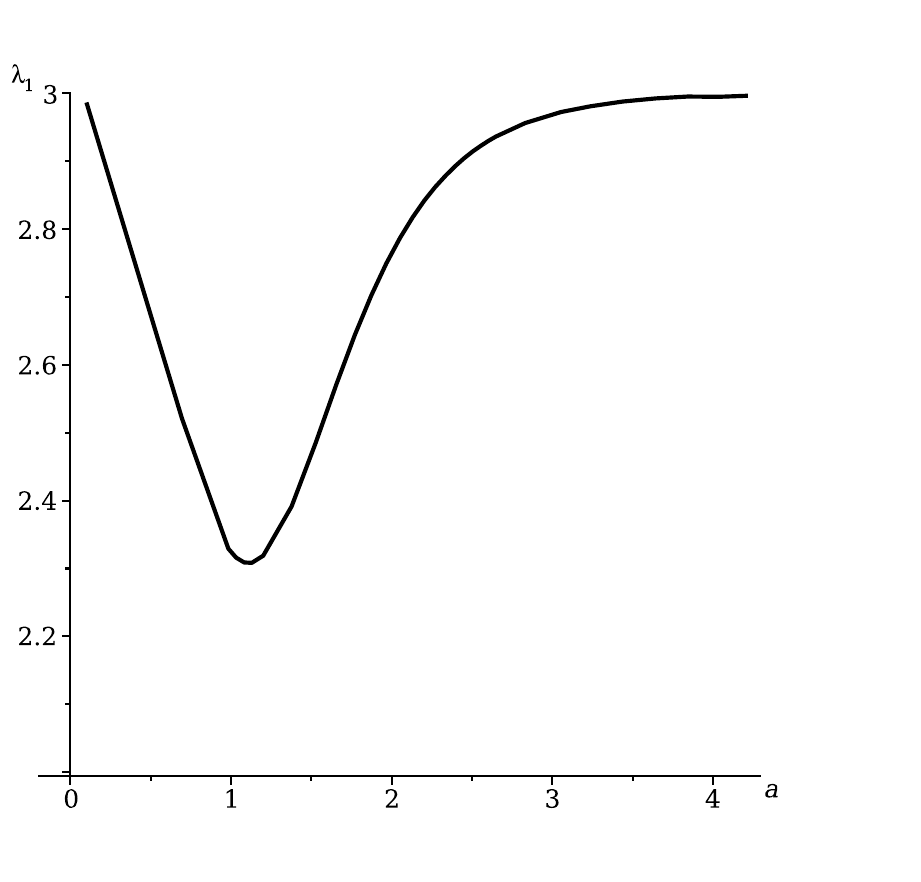}
\includegraphics[width=0.53\textwidth,trim=2cm 12cm 2cm 1cm, clip]{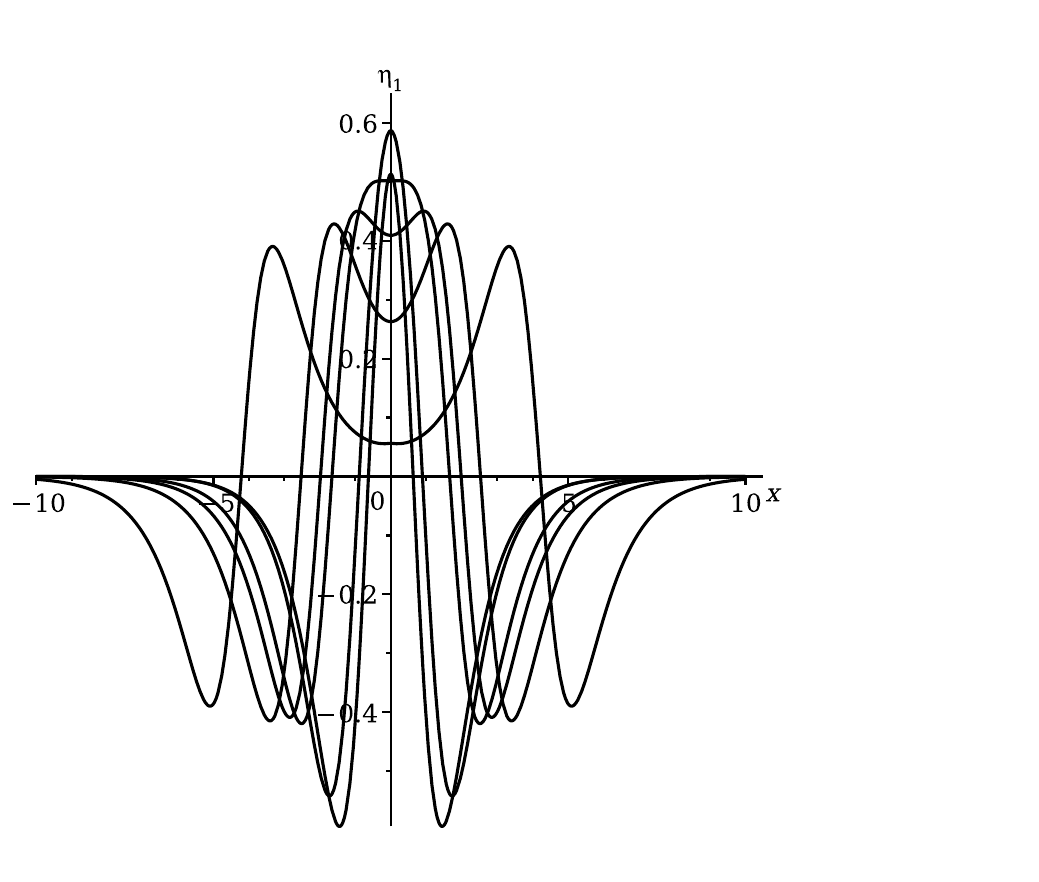}
\caption{(Left)\quad $\lambda_{1}$;\quad
(Right)\quad $\eta_{1}(x)$, normalized in $L^2$, 
for $a$ = 0.695, 1.082, 1.771, 2.052, 2.555, 4.218}
\label{fig:2node.eigenvalue.eigenfunction}
\end{figure}

The eigenvalue equation \eqref{eigenvalue.deteqn} can be investigated numerically
to look for other solutions, which would describe symmetric eigenfunctions. 
We find that it possesses one positive solution, which we denote as $\lambda_{1}$.
Sturm's oscillation theorem implies that the corresponding symmetric eigenfunction,
$\eta_{1}(x) = \tilde \eta_{1}(y)$, 
has two nodes.
It is called an internal vibrational (or shape) mode.
As it has a positive eigenvalue, this mode is linearly stable. 
Fig.~\ref{fig:2node.eigenvalue.eigenfunction} shows a plot of
the eigenvalue and corresponding eigenfunction.

\begin{figure}
\includegraphics[width=0.45\textwidth,trim=2cm 12cm 4cm 1cm, clip]{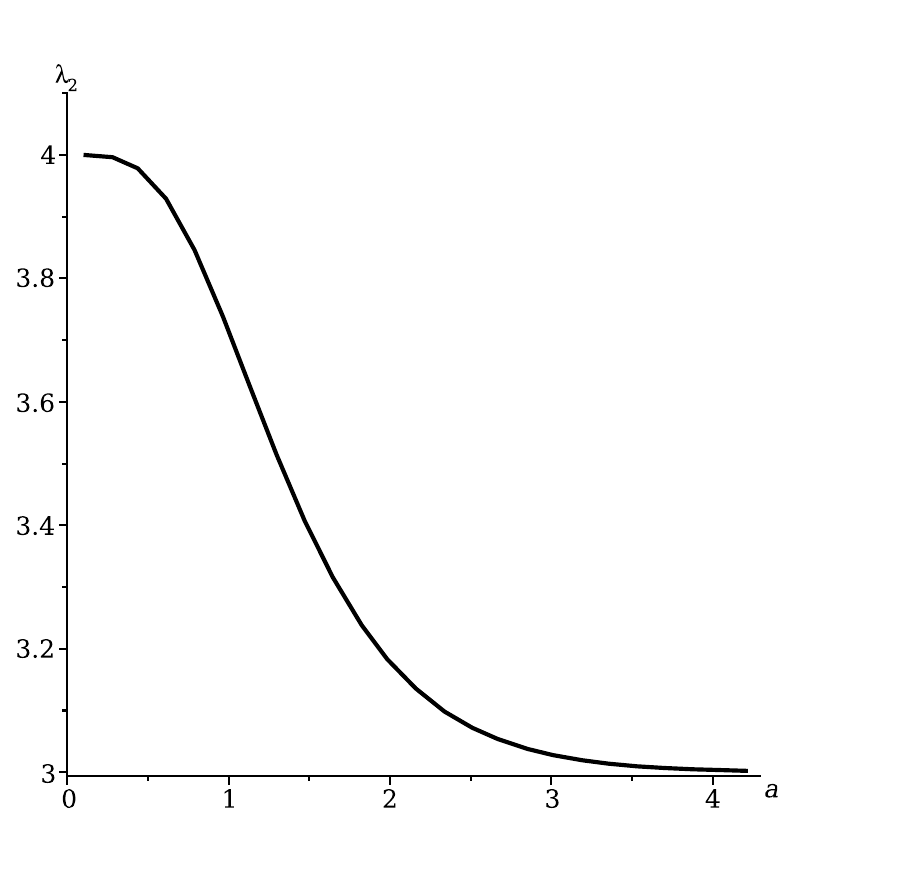}
\includegraphics[width=0.54\textwidth,trim=2cm 12cm 2cm 1cm, clip]{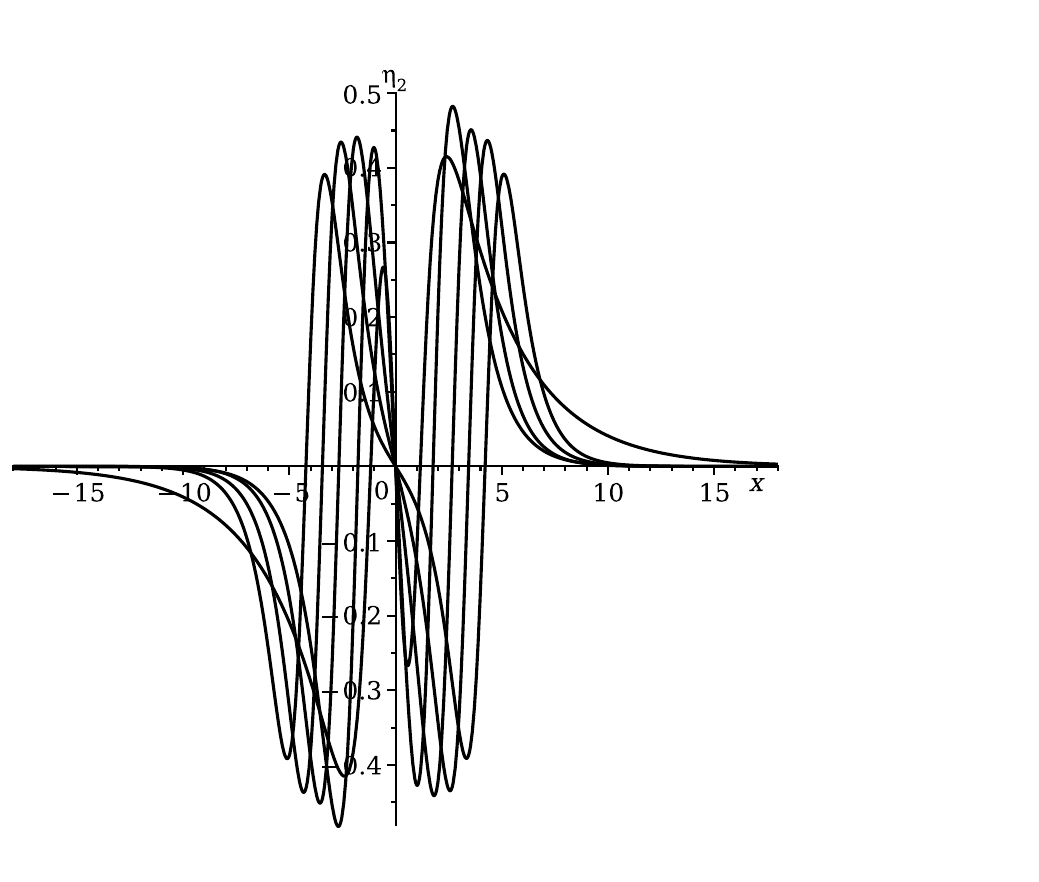}
\caption{(Left)\quad $\lambda_{2}$;\quad
(Right)\quad $\eta_{2}(x)$, normalized in $L^2$, 
for $a$ = 0.695, 1.656, 2.649, 3.438, 4.218}
\label{fig:3node.eigenvalue.eigenfunction}
\end{figure}

Since the continuous spectrum starts at $\lambda = 4$,
what remains to be determined is if there is another positive eigenvalue $\lambda_{2}>\lambda_{1}$
corresponding to an eigenfunction that has three nodes. 
The determining condition for the eigenvalue is simply that
$y^{\sqrt{1-\lambda/4}}(\sinh(a)^2 y + 1)^3 H(y)$ must vanish at $y=1$. 
We find that it does have a solution, which is plotted in Fig.~\ref{fig:3node.eigenvalue.eigenfunction}. 
The corresponding eigenfunction $\eta_{2}(x) = \tilde \eta_{2}(y)$ is also plotted. 
This represents an additional internal vibrational mode, which is linearly stable. 

This completes the determination of the spectrum.

\section{Approximate analytical formula for the unstable mode}
\label{sec:approx}

No closed form expression is known for the local Heun functions 
appearing in the ground state eigenvalue equation \eqref{eigenvalue.deteqn}
and in the ground state eigenfunction \eqref{-1.eigenfunct.x}. 
However, the eigenvalue $\lambda_{-1}$ can be shown to satisfy 
an explicit continued-fraction equation arising from expressions \eqref{H(y)}--\eqref{H(y).alt}
as follows. 

Since the eigenvalue equation is equivalent to the condition $C_2=0$, 
expressions \eqref{H(y)}--\eqref{H(y).alt} with $\lambda=\lambda_{-1}$ 
thereby show that the functions 
$H\ell(p,q;\alpha,\beta,\gamma,\delta;y)|_{\lambda=\lambda_{-1}}$
and $H\ell(1-p,\alpha\beta - q;\alpha,\beta,\delta,\gamma;1-y)|_{\lambda=\lambda_{-1}}$
are analytic at both $y=0$ and $y=1$. 
As known from general results stated in \Ref{Ron-book},
this analyticity is equivalent to a condition on the coefficients 
in the corresponding series expansions. 
We will use the series for 
$H\ell(1-p,\alpha\beta - q;\alpha,\beta,\delta,\gamma;1-y)|_{\lambda=\lambda_{-1}}$
because, for any value of $a$, 
its convergence interval includes both points $y=0$ and $y=1$. 
The series coefficients, $c_i$, satisfy a three term recurrence relation
$R_j c_{j+1} - (Q_j +\alpha\beta- q) c_j + P_j c_{j-1} =0$, $j=1,2,\ldots$, 
with $c_0=0$ and $c_1=1$, 
where 
\begin{equation}\label{Heun.series.RQP}
R_j = (1-p)(j+1)(j+\delta),
\quad
Q_j = j( (j-1+\delta)(2-p) + (1-p)\gamma + \epsilon),
\quad
P_j = (j-1+\alpha)(j-1+\beta) . 
\end{equation}
The condition of analyticity at the two points $y=0$ and $y=1$ is given by 
the continued fraction equation \cite{Ron-book}
\begin{equation}\label{q.eqn}
q = 
\frac{R_0 P_1}{Q_1 + q -} 
\cdots \frac{R_j P_{j+1}}{Q_{j+1} + q - } \cdots
. 
\end{equation}
This becomes an explicit equation 
once the parameters \eqref{Heun.ode.params} have been substituted. 
It can be used to get an approximation for $\lambda = \lambda_{-1}$ 
by truncating the expansion. 
Truncation at $j=0$ gives a quartic equation in terms of
$\mu=\sqrt{4-\lambda_{-1}}$, 
\begin{equation}\label{mu.eqn}
\begin{aligned}
& \cosh(a)^2 \mu^4   + 6 \cosh(a)^2 \mu^3   + (47\cosh(a)^2 - 60) \mu^2 
\\&\qquad
+ (42\cosh(a)^2 - 108) \mu + 72 (2\cosh(a)^2 - 12 + 11 \sech(a)^2) =0, 
\end{aligned}
\end{equation}
which yields an explicit approximate formula for the eigenvalue $\lambda_{-1}$. 
Compared to the actual numerical value, 
this approximation turns out to be good only in a small parameter range 
$a \lesssim 1$. 

An approximate asymptotic formula for both the eigenvalue and eigenfunction
can be obtained near the end points $a\to 0$ and $a\to \infty$ of the parameter range. 
From Fig.~\ref{fig:neg.eigenvalue}, note that
$\lambda_{-1}\to -5$ as $a\to 0$,
and $\lambda_{-1}\to 0$ as $a\to \infty$.

For $a$ near $0$,
the approximate equation \eqref{mu.eqn} is applicable and yields
$\mu \simeq 3 - \tfrac{8}{7} a^2$ to lowest order.
This gives
\begin{equation}
\lambda_{-1} \simeq -5 +\tfrac{48}{7} a^2 . 
\end{equation}
Substituting this expression into the eigenfunction equation \eqref{lump.eigenfunct.eqn.y}
leads to the solution 
\begin{equation}
\tilde\eta_{-1}(y) =
y^{\frac{3}{2} -\frac{4}{7}a^2} 
H\ell_{\text c}(0, 3 - \tfrac{8}{7} a^2, -1/2, 9 a^2,- \tfrac{54}{7} a^2 - 1/2, y)
\end{equation}
where $H\ell_\text{c}$ denotes the confluent local Heun function \cite{Ron-book}.
Plotting this function shows that it can be well approximated by
$H\ell_{\text c}(0, 3 - \tfrac{8}{7} a^2, -1/2, 9 a^2,- \tfrac{54}{7} a^2 - 1/2, y)
\simeq
1-2a^2 y$.
The transformation \eqref{chgvars.y}
then yields 
$\eta_{-1}(x) \simeq (1 + 2 a^2 \sech(x)^2)\sech(x)^{3 + \frac{8}{7} a^2}$.
After expanding 
$\sech(x)^{a^2} \simeq 1 + a^2\ln\sech(x)$,
followed by normalizing so that $\eta_{-1}(0)=1$,
we finally obtain the approximate ground state eigenfunction
\begin{equation}
\eta_{-1}(x)   \simeq
\sech(x)^3\big( 1 + \tfrac{2}{7} a^2 \big(7\tanh(x)^2 - 4\ln(\sech(x))\big) \big)
\end{equation}
when $a$ is near $0$. 

For large $a$,
there is no a priori approximation available for the eigenvalue. 
Instead we start from the exact eigenfunction \eqref{-1.eigenfunct.y},
put $1-\lambda_{-1}/4= (1+\mu)^2$ with $\mu$ assumed to be small, 
and approximate the parameters \eqref{Heun.ode.params}:
\begin{equation}
p \simeq  -4 e^{-2a}, 
\quad
q \simeq 15 + 6\mu + 6 e^{-2a},
\quad
\alpha \simeq 4 + \mu,
\quad
\beta \simeq \tfrac{9}{2} + \mu,
\quad
\gamma \simeq 3 + 2\mu,
\quad 
\delta = 1/2 . 
\end{equation}
We next examine the eigenvalue equation \eqref{eigenvalue.deteqn} numerically
and find that it yields 
$\mu \simeq 12 e^{-4a}$.
The eigenvalue is thus given by
\begin{equation}
\lambda_{-1} \simeq -96 e^{-4a}. 
\end{equation}
This gives the following approximation for the eigenfunction
\begin{equation}
\tilde\eta_{-1}(y) \simeq 
\tfrac{1}{8} y^{1+12 e^{-4a}} (4+y e^{2a})^3 H(y)
\end{equation}
where
\begin{equation}
 H(y) = H\ell(-4 e^{-2a}, 6 e^{-2a} + 15, 12 e^{-4a} + 4, \tfrac{9}{2} + 12 e^{-4a}, 24 e^{-4a}+ 3, 1/2, y) . 
\end{equation}
Now we would like to approximate this function in terms of elementary functions. 
Plotting it, we see that
$H(y) \simeq \big(2-y\big)^4/\big(2(1-y)+ e^{2a}y\big)^4$
yields a good approximation.
We finally expand $\sech(x)^{e^{-4a}} \simeq 1 + e^{-4a}\ln\sech(x)$,
and normalize $\eta_{-1}(0)=1$,
which yields the approximate ground state eigenfunction 
\begin{equation}
\eta_{-1}(x) \simeq
\big(2 -\sech(x)^2\big)^4 \big(4 e^{-2a}\sinh(x)^2 + 1\big)^3 \big(1 + 24 e^{-4a} \ln\sech(x)\big) /\big(2 e^{-2a}\sinh(x)^2 + 1\big)^4
\end{equation}
when $a$ is large.

\subsection{Time scale}

\begin{figure}
\includegraphics[width=0.60\textwidth,trim=2cm 12cm 2cm 1cm, clip]{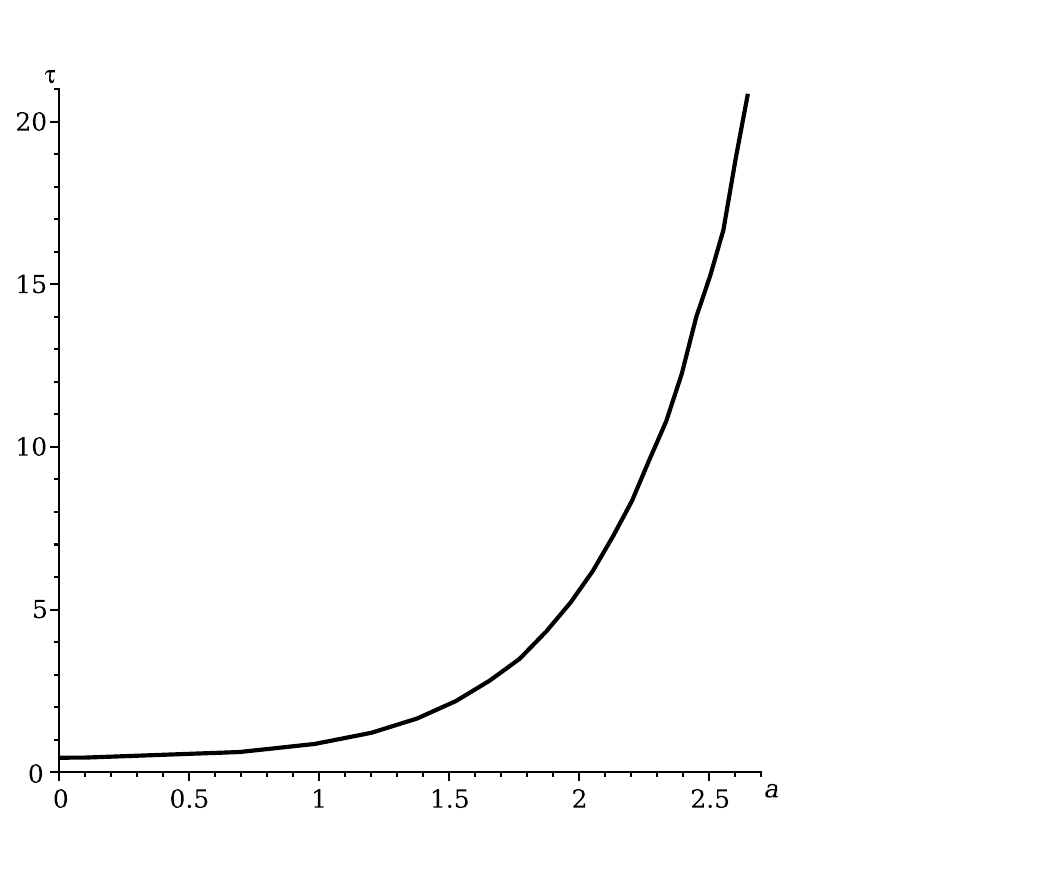}
\caption{Lifetime $\tau$}
\label{fig:lifetime}
\end{figure}

The time scale of the growing perturbation mode is given by 
\begin{equation}
\tau = 1/\sqrt{|\lambda_{-1}|} 
\end{equation}
which is plotted in Fig.~\ref{fig:lifetime}.
When the sphaleron is perturbed and becomes unstable,
this time scale represents the approximate lifetime of the sphaleron
before the change due to instability becomes large. 

As $a\to0$, we have
$\tau \simeq \tfrac{1}{\sqrt{5}}(1 +\tfrac{24}{35} a^2) \to 0.4472$,
while $\tau \simeq \tfrac{1}{4\sqrt{6}} e^{2a} \to \infty$
as $a\to \infty$.

\section{Concluding remarks}
\label{sec:remarks}

The spectral problem for sphalerons has been studied
for a general quartic Klein-Gordon model with a false vacuum. 
Several new results are obtained. 

\begin{itemize}
\item
The eigenfunction and eigenvalues are found analytically in terms of local Heun functions. 
\item
In addition to an unstable ground-state mode, 
and the translation (zero) mode, 
the problem is found to have two internal vibrational modes. 
\item 
Approximate explicit expressions in terms of elementary functions are developed
for the ground state eigenvalue and eigenfunction. 
\end{itemize}

These rigorous analytical results provide a foundation
for subsequent investigations of perturbed sphaleron evolution 
in nonlinear Klein-Gordon models \cite{Anc.Saa,Saa.Amu.Anc}.

\appendix

\section{Conservation laws}\label{app:conslaws}

Continuous symmetries of the action principle \eqref{KG.action}
give rise to conservation laws via Noether's theorem.
For a general potential, the continuous symmetries consist of 
time-translations, space-translations, and Lorentz boosts.
These symmetries are inherited by the equation of motion \eqref{KG.eqn}.

Time-translations yield energy 
\begin{equation}\label{ener}
E[\phi] = \int_{-\infty}^{\infty} \Big( \tfrac{1}{2} \phi_t^2 + \tfrac{1}{2} \phi_x^2 + V(\phi) \Big)\, dx . 
\end{equation}
Space-translations yield linear momentum
\begin{equation}\label{mom}
P[\phi] = \int_{-\infty}^{\infty}\Big( {-}\phi_t \phi_x\Big) \,dx . 
\end{equation}
Lorentz boosts yield boost momentum
\begin{equation}\label{boostmom}
J[\phi] = \int_{-\infty}^{\infty} \Big( t \phi_t \phi_x + x\big(\tfrac{1}{2} \phi_t^2 + \frac{1}{2} \phi_x^2 + V(\phi) \big) \Big)\,dx . 
\end{equation}
These three integrals are conserved for all solutions $\phi(x,t)$
with sufficient asymptotic decay as $x\to\pm\infty$.

The boost momentum can be expressed as
\begin{equation}
J[\phi] = \chi[\phi;t] E[\phi] -t P[\phi]
\end{equation}
where
\begin{equation}
\chi[\phi;t] = \frac{1}{E[\phi]} \int_{-\infty}^{\infty} x\Big(\tfrac{1}{2} \phi_t^2 + \tfrac{1}{2} \phi_x^2 + V(\phi) \Big)\,dx
\end{equation}
defines the center of energy. 
Conservation of $E$ and $P$ thereby implies $\frac{d}{dt}\chi = P/E$,
showing that the center of energy moves at constant speed.

\section*{Acknowledgements}

The author is supported by an NSERC Discovery grant. 

Danial Saadatmand is thanked for inspiring this study,
and Alexandr Chernyavskiy is thanked for valuable discussions. 

Data sharing not applicable to this article as no datasets were generated or analysed during the current study.

The author has no conflict of interest to declare.

\end{document}